\documentclass[aps,preprint,pre,superscriptaddress,runinaddress]{revtex4} 
\usepackage{amssymb}

\usepackage{graphicx}

\begin{document}
%%    The information for the title page will be placed between
%%    \begin{document} and \maketitle. The order of most entries
%%    is determined by the class file and can not be changed by
%%    rearranging them. The maketitle command follows after the
%%    abstract.
%%
%%    Most of the following commands will be completed by the publisher.
%%
%%    The copyrightyear is defined in the .clo file as the first argument
%%    of the copyrightinfo command. If the copyrightyear differs from that
%%    value it might be adjusted by the following definition:
%%
%% \renewcommand{\copyrightyear}{2007}% uncomment to change the copyrightyear.
%%
%\DOIsuffix{theDOIsuffix}
%%
%% issueinfo for the header line
%\Volume{46}
%\Month{01}
%\Year{2007}
%%
%%    First and last pagenumber of the article. If the option
%%    'autolastpage' is set (default) the second argument may be left empty.
%\pagespan{1}{}
%%
%%    Dates will be filled in by the publisher. The 'reviseddate' and
%%    'dateposted' (Published online) entry may be left empty.
%\Receiveddate{XXXX}
%\Reviseddate{XXXX}
%\Accepteddate{XXXX}
%\Dateposted{XXXX}
%%

%% \pretitle{Editor's Choice}

%% We have a short and a long form for the title. The short form
%% (optional argument) goes into the running head.

\title{Charge Correlations in a Harmonic Trap}

%% Please do not enter footnotes or \inst{}-notes into the optional
%% argument of the author command. The optional argument will go into
%% the header.  If there is only one address the marker \inst{x} may be
%% omitted.

%% Information for the first author.
%\author[J. Wrighton]{Jeffrey Wrighton\inst{1,}\footnote{Corresponding author\quad
%E-mail:~\textsf{wrighton@phys.ufl.edu}}}
%\address[\inst{1}]{Department of Physics, University of Florida, Gainesville, FL 32611}
%\author[H. K\"ahlert]{Hanno K\"ahlert\inst{2}}
%\address[\inst{2}]{Institut f\"{u}r Theoretische Physik und Astrophysik, Christian-Albrechts
%Universit\"{a}t zu Kiel, 24098 Kiel, Germany}
%\author[T. Ott]{Torben Ott\inst{2}}
%\author[P. Ludwig]{Patrick Ludwig\inst{2}}
%\author[H. Thomsen]{Hauke Thomsen\inst{2}}
%\author[J. Dufty]{James Dufty\inst{1}}
%\author[M. Bonitz]{Michael Bonitz\inst{2}}

\author{Jeffrey Wrighton}
\affiliation{Department of Physics, University of Florida, Gainesville, FL 32611}
\author{Hanno K\"ahlert}
\author{Torben Ott}
\author{Patrick Ludwig}
\author{Hauke Thomsen}
\affiliation{Institut f\"{u}r Theoretische Physik und Astrophysik, Christian-Albrechts
Universit\"{a}t zu Kiel, 24098 Kiel, Germany}
\author{James Dufty}
\affiliation{Department of Physics, University of Florida, Gainesville, FL 32611}
\author{Michael Bonitz}
\affiliation{Institut f\"{u}r Theoretische Physik und Astrophysik, Christian-Albrechts
Universit\"{a}t zu Kiel, 24098 Kiel, Germany}

%%\author[F. Author]{First Author\inst{1,}%
%%  \footnote{Corresponding author\quad E-mail:~\textsf{x.y@xxx.yyy.zz},
%%            Phone: +00\,999\,999\,999,
%%            Fax: +00\,999\,999\,999}}
%%\address[\inst{1}]{First address}
%%
%%    Information for the second author
%%\author[S. Author]{Second Author\inst{1,2,}\footnote{Second author footnote.}}
%%\address[\inst{2}]{Second address}
%%
%%    Information for the third author
%%\author[T. Author]{Third Author\inst{2,}\footnote{Third author footnote.}}
%%
%%    \dedicatory{This is a dedicatory.}
\begin{abstract}
  A system of $N$ classical Coulomb charges trapped in a harmonic
potential displays shell structure and orientational ordering. The
local density profile is well understood from theory, simulation,
and experiment. Here, pair correlations are considered for this
highly inhomogeneous system for both the fluid and ordered states.
In the former, it is noted that there is a close relationship to
pair correlations in the uniform OCP. For the ordered state, it is
shown that the disordered ``tiling" is closely related to the ground
state Thomson sites for a single sphere.
\end{abstract}
%% maketitle must follow the abstract.
\maketitle                   % Produces the title.

%% If there is not enough space inside the running head
%% for all authors including the title you may provide
%% the leftmark in one of the following three forms:

%% \renewcommand{\leftmark}
%% {First Author: A Short Title}

%% \renewcommand{\leftmark}
%% {First Author and Second Author: A Short Title}

%% \renewcommand{\leftmark}
%% {First Author et al.: A Short Title}

%% \tableofcontents  % Produces the table of contents.
\section{Introduction}

Harmonically trapped Coulomb charges form a shell structure in their
local density profile at sufficiently strong coupling. An ordering
of the particles within the shells occurs at still stronger
coupling. This behavior is now well understood at strong coupling
from both experiment and
simulation \cite%
{Pohl2004,Arp,Ludwig2005,bonitz2006,Golubnychiy2006,Baumgartner} 
and from theory (shell models)
\cite{Baumgartner,Kraeft06,Cioslowski08}. Recently, Monte Carlo
simulation and density functional theory have been applied across
the entire fluid phase \cite{Wrighton09,Wrighton10}. The theoretical
approaches make clear the essential role of pair correlations in the
formation of shells. Somewhat surprisingly, it was found that good
shell structure could be obtained using the pair correlations for a
uniform one component plasma (OCP) as an approximation to the true
correlations of the highly non-uniform trap. The objective here is
to report some initial studies of pair correlations among charges in
a trap under conditions of strong shell structure in both the fluid
and ordered states. The primary observations are:

\begin{itemize}
\item The distribution of pairs in the trap in their fluid state is well
represented by that for the three dimensional OCP for particle number $%
N\gtrsim 50$, even at strong coupling where shell structure is
prominent.

\item Angular correlations for particles within a given shell in the fluid
phase are similar to those obtained from the uniform three
dimensional OCP if the pair distance of the latter is reinterpreted
as the chord length for two points in the shell.

\item Angular correlations for charges confined to a single sphere in the
fluid phase are almost identical to those for the two dimensional
OCP with the pair distance interpreted as the chord length.

\item Angular correlations in the ordered phase can be represented by the
solutions to the Thomson problem \cite{Thomson} broadened by thermal
disorder.
\end{itemize}

\section{Pair correlations - fluid phase}
\begin{figure}
\includegraphics[width=68mm]{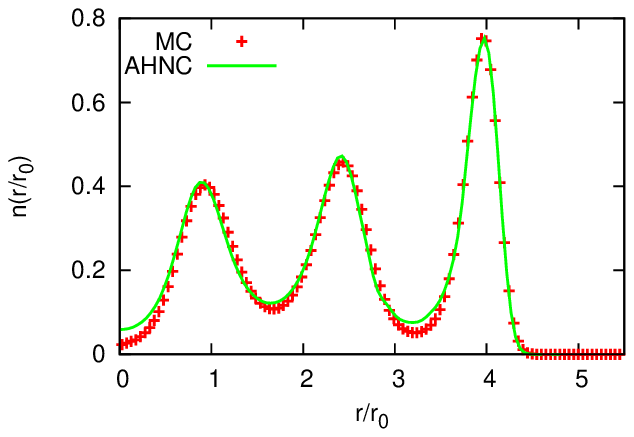}~a)
\hfil
\includegraphics[width=68mm]{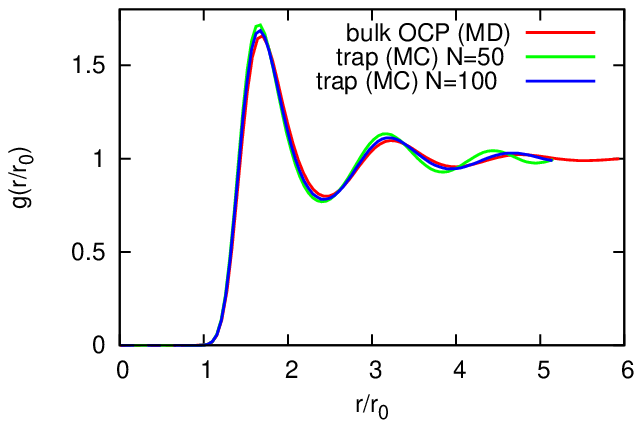}~b)
\caption{Density profile for $N=100$ charges in trap showing shell structure ($\Gamma=50$)
(\textbf{a}). Comparison of distribution of pairs for particles in
trap with that for particles in bulk OCP ($\Gamma=50$) (\textbf{b}).}
\label{fig1}
\end{figure}
\label{sec2} Consider a one component system of Coulomb charges $q$. 
The two cases of interest here
are a system of charges in a harmonic trap, and the same system of
charges in a uniform neutralizing background (OCP). Only equilibrium
correlations are considered so the
statistical properties are determined from the potential energies%
\begin{equation}
U_{T}\left( \left\{ r_{i}\right\} \right)
=\sum_{i=1}^{N}\frac{1}{2}m\omega
^{2}r_{i}^{2}+\frac{1}{2}\sum_{i\neq j}^{N}\frac{q^{2}}{\left\vert \mathbf{r}%
_{i}-\mathbf{r}_{j}\right\vert },  \label{2.1}
\end{equation}%
\begin{equation}
U_{OCP}\left( \left\{ r_{i}\right\} \right) =\frac{1}{2}\sum_{i\neq j}^{N}%
\frac{q^{2}}{\left\vert \mathbf{r}_{i}-\mathbf{r}_{j}\right\vert }%
-\sum_{i=1}^{N}n\int d\mathbf{r}^{\prime }\frac{q^{2}}{\left\vert \mathbf{r}%
_{i}-\mathbf{r}^{\prime }\right\vert }+n^{2}\frac{1}{2}\int d\mathbf{r}%
^{\prime }d\mathbf{r}^{\prime \prime }\frac{q^{2}}{\left\vert \mathbf{r}%
^{\prime }-\mathbf{r}^{\prime \prime }\right\vert }.  \label{2.2}
\end{equation}%
Here $n=N/V$, where the volume is taken to be spherical with radius
$R$.

Usually, the OCP is considered in the thermodynamic limit
($N,V\rightarrow \infty ,n=N/V=$constant) where it is spatially
uniform. This will be referred to as the bulk OCP. However,  
the charges in the trap and those of the OCP
constitute the same system for finite $N$ \cite{Dubin1999}, which is shown by performing the integrals in
(\ref{2.2}) to give
\begin{equation}
U_{OCP}=\frac{1}{2}\sum_{i\neq j}^{N}\frac{q^{2}}{\left\vert \mathbf{r}_{i}-%
\mathbf{r}_{j}\right\vert }+\frac{2\pi q^{2}}{3}n\sum_{i=1}^{N}r_{i}^{2}+%
\frac{8\pi q^{2}}{5}nNR^{2}.  \label{2.5}
\end{equation}%
Thus $U_{OCP}$ is equivalent to $U_{T}$, up to a
constant which does not contribute to the equilibrium ensemble.
Both systems are self-bound at a maximum radius $R_{0}$ (when
$R>R_{0}$).  
Force equilibrium
on a particle  for the trap $m\omega
^{2}R_{0}=q^{2}N/R_{0}^{2}$ and similarly for the OCP $4\pi
q^{2}nR_{0}/3=q^{2}N/R_{0}^{2}$ determine $R_{0}$. Thus for the same $N$ in the
trap and OCP, their volume and average density are the same. In particular $%
n_{OCP}=n_{T}=3m\omega ^{2}/4\pi q^{2}$ and  $R_{0}^{3}=q^{2}N/m\omega ^{2}$%
. This also implies $\omega _{p}^{2}=3\omega ^{2}$, where
$\omega _{p}^{2}=4\pi n_{OCP}q^{2}/m$ defines the plasma frequency.
In this context, the bulk OCP corresponds to a harmonic trap with
infinite filling, and the shell structure for particles in a trap
can be understood as finite volume effects for the OCP. The
terminology here will  be OCP for finite $N$ and bulk OCP for
the thermodynamic limit. Simulations using periodic boundary
conditions describe the bulk OCP.

In reference \cite{Wrighton09} a theory for the density profile in
the trap was developed in terms of the pair correlations within the
trap. For practical purposes it was found that the corresponding
bulk OCP pair correlations could be used, giving an accurate
approximation for the shell structure. This is illustrated in Fig.
\ref{fig1}a for $N=100$, and coupling strength $\Gamma \equiv
\beta q^{2}/r_{0}=50$ (where $r_{0}$ is the mean distance for a pair
defined by $4\pi nr_{0}^{3}/3=1$ and $n=N/V$ is the density).
Results from the adjusted hypernetted chain theory (AHNC) and Monte Carlo are
given, showing their good agreement. In spite of the above
equivalence of the OCP and trap for finite $N$, it is surprising
that the correlations for the uniform bulk OCP could be the same as
those for the trap with strong shell structure. Nevertheless, this
is the case as shown in Fig. \ref{fig1}b. The agreement is quite reasonable
for $N=50$ and improves with increasing $N$. This plot gives the
distribution of pairs within the trap from Monte Carlo simulation,
without reference to where the center of mass of the pair is \
located. Thus it is not the pair correlation function defined
relative to the center of the trap, which would indeed reflect its
spatial inhomogeneity. In Fig. \ref{fig1}b the contributions
to a given $r$ come from all pairs at that distance anywhere within
the trap. The bulk three dimensional OCP pair correlation function
is determined from molecular dynamics simulation.
%The \LaTeX{} code for Figs. \ref{fig:5}a and b is
%\begin{verbatim}
%\begin{figure}
%\includegraphics[width=68mm,height=25mm]{empty}~a)
%\hfil
%\includegraphics[width=68mm,height=25mm]{empty}~b)
%\caption{Two figures with one number. The figures are referred to as
%\textbf{a} and \textbf{b}.} \label{fig:5}
%\end{figure}
%\end{verbatim}

This agreement is possible because the property being calculated
depends only on the relative separation of pairs, a translationally
invariant property. The trap Hamiltonian can be expressed in terms
of its center of mass   and relative coordinates by a
canonical transformation. For averages of properties depending only on
pair separation the center of mass coordinate can be integrated out
leaving a translationally invariant potential. For sufficiently
large $N$ (e.g., $N=50$) most  pairs are away from the 
trap surface and the pair distribution behaves like that for the bulk
OCP. A detailed demonstration of this will be given elsewhere.

\section{Angular correlations - fluid phase}

Next consider the particles within a chosen shell, defined by the
domain between the minima on either side of a peak in the radial density
profile (see Fig. {\ref{fig1}a). An initial particle is chosen
and the   number of particles at an angle }$\theta $ relative
to the first is calculated. In the fluid phase there is rotational
symmetry about the line from the origin to the first particle, so
only one angle is required for this correlation function. In
practice the results obtained by Monte Carlo simulation are an
average over the radial annulus of the shell for both members of a
pair.

\begin{figure}
\includegraphics[width=68mm]{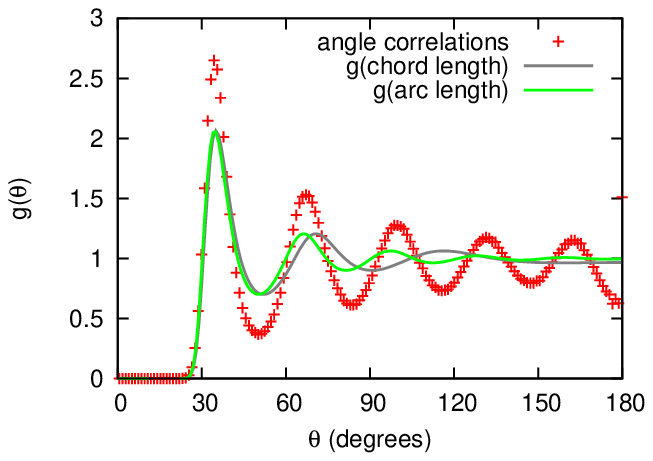}~a)
\hfil
\includegraphics[width=68mm]{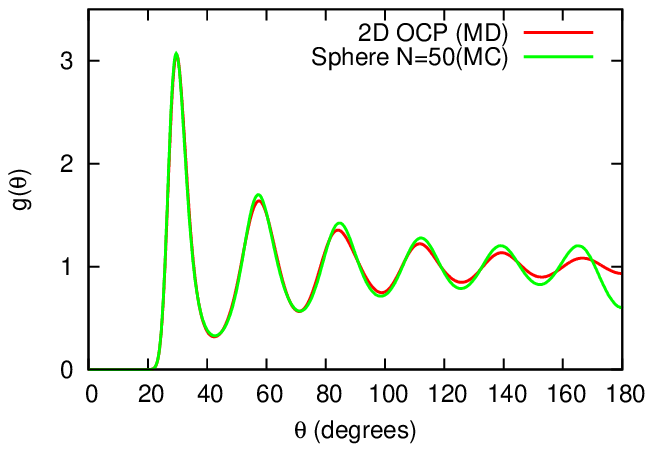}~b)
\caption{Angular correlations for particles in the outer shell
compared to those for a 3d bulk OCP ($N=44$, $\Gamma=100$) (\textbf{a}). Pair correlation
function for particles confined to a single sphere ($N=50$) compared to that
for a 2d bulk OCP with distance interpreted as arc length ($\Gamma=126$) (\textbf{b}).}
\label{fig2}
\end{figure}

Fig. {\ref{fig2}a shows the angular correlation function from
molecular dynamics simulation of }$N=44,\Gamma =100$, for which there
are two shells with $8$ and $36$ particles. The correlation function
is for the larger outer shell. Also shown are two results from the
same molecular dynamics simulation of the three dimensional OCP at
the same value of $\Gamma $, but with two interpretations for the
pair separation. In one case, the usual Euclidean distance between
particles is chosen, i.e. the chord length. In the second case the
argument of the bulk OCP pair correlation function is interpreted as
the larger arc length. The figure shows a definite improvement in
relative agreement with the trap correlations in the second case.
Absolute agreement is not expected since the trap data is averaged
over the annulus whereas the OCP data is calculated for points at
the maximum of the outer shell only. In this context the location of
the peaks is the relevant test for qualitative comparison.

The complication of a finite shell width  can be
mitigated by increasing the coupling constant $\Gamma $, leading to
a sharpening of the shells into smaller annulae. This is
limited since ordering within the shells occurs at some maximum
value of $\Gamma $ for the fluid state. The correlations are then
very different (see below) and not related
to those of the bulk OCP. To explore the interesting relationship of Fig. %
{\ref{fig2}a under more controlled conditions, consider the
idealized case of charges confined to a single spherical shell of
zero width. The correlations are now constrained to a two
dimensional surface and the appropriate comparison is with
correlations in the two dimensional OCP. This is demonstrated in
Fig. \ref{fig2}b where the angular correlations for particles
constrained to the sphere at $\Gamma =126$ are compared
to those for the two dimensional (Coulomb potential) OCP pair
correlation function \emph{with the distance reinterpreted as the arc length}%
. The remarkable agreement suggests a mathematical relationship
between the two quite different systems. Why should the two
 systems find agreement when their geometry (metric) is
adjusted? This will be discussed elsewhere.

\section{Angular correlations - ordered phase}

At sufficiently large $\Gamma $ rotational symmetry is broken and
the particles within each shell become ordered. The ground state
configuration for $\Gamma \rightarrow \infty $ is well studied by
simulation and theory. In particular, a shell model using the
correlation energy for the Thomson sites (minimum energy
configuration for charges on a   sphere) gives an excellent
description of the trap ground state energy \cite{Cioslowski08}. In
fact, the ground state positions for a given shell from quenched molecular simulation 
 are very close to the
Thomson sites on a sphere of the same size and particle
number. Due to the spherical geometry, the ordering is not regular
in general and depends on the particle number. It is tempting to
consider the Thomson sites as the analogue of a fundamental lattice
for these spherical crystals.

\begin{figure}[<float>]
%\sidecaption
\includegraphics[width=68mm]{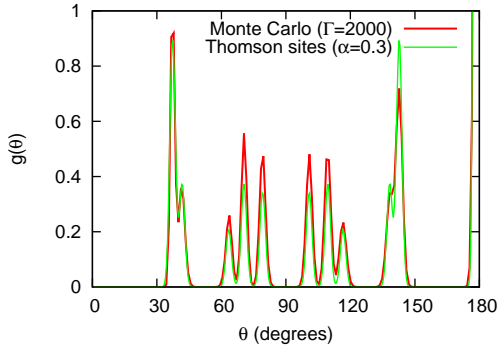}%
\caption{Comparison of the angular correlations among particles in
the outer shell with those for the thermally broadened Thomson
sites ($N=38$).}
\label{fig3} % Give a unique label
\end{figure}

To test this picture  the angular
correlations for particles of one shell of a trap with those for the
corresponding Thomson sites are compared.  
At finite $\Gamma$, the Thomson site charges have 
kinetic energy and are represented by
\begin{equation}
f(\theta )=\sqrt{\frac{\alpha }{\pi }}e^{-\alpha \left( \theta
-\theta _{0}\right) ^{2}},
\end{equation}%
where $\theta _{0}$ are the Thomson sites and $\alpha $ is a function of $%
\Gamma $.  Configuration data for the Thomson sites can be found at \cite{thomapp}. 
Fig. \ref{fig3} shows this comparison for a trap
with $38$ particles ($32$ in the outer shell) at $\Gamma =2000$. The
corresponding Thomson correlations are shown for $\alpha =0.3$. The
very good agreement provides initial support for this picture of
Thomson sites as a fundamental spherical ``lattice". A detailed test
requires further analysis to study how structure appears as
$\Gamma $ is increased, other values of $N$ for which very different
order occurs, and a comparison with configurations for nearby
metastable states. This will be reported elsewhere.

\section{Acknowledgements}
This work is supported by the Deutsche Forschungsgemeinschaft via
SFB-TR 24, and by the NSF/DOE Partnership in Basic Plasma Science
and Engineering under the Department of Energy award
DE-FG02-07ER54946.


\begin{thebibliography}{99}
\bibitem{Pohl2004} T. Pohl, T. Pattard, and J.M. Rost, Phys. Rev. Lett.
\textbf{92}, 155003 (2004).

\bibitem{Arp} O. Arp, D. Block, A. Piel, and A. Melzer, Phys. Rev. Lett.
\textbf{93}, 165004 (2004); O. Arp, D. Block, M. Bonitz, H. Fehske,
V. Golubnychiy, S. Kosse, P. Ludwig, A. Melzer, and A. Piel, J.
Phys. Conf. Series \textbf{11}, 234 (2005).

\bibitem{Ludwig2005} P. Ludwig, S. Kosse, and M. Bonitz, Phys. Rev. E 71,
046403 (2005).

\bibitem{bonitz2006} M. Bonitz, D. Block, O. Arp, V. Golubnychiy, H.
Baumgartner, P. Ludwig, A. Piel, and A. Filinov, Phys. Rev. Lett. \textbf{96}%
, 075001 (2006).

\bibitem{Golubnychiy2006} V. Golubnychiy, H. Baumgartner, M. Bonitz, A.
Filinov, and H. Fehske, J. Phys. A: Math. Gen. \textbf{39}, 4527
(2006).

\bibitem{Baumgartner} H. Baumgartner, H. K\"{a}hlert, V. Golubnychiy, C.
Henning, S. K\"{a}ding, A. Melzer, and M. Bonitz, Contrib. Plasma
Phys. \textbf{47}, 281 (2007); H. Baumgartner, D. Asmus, V.
Golubnychiy, P. Ludwig, H. K\"{a}hlert, and M. Bonitz, New Journal
of Physics \textbf{10}, 093019 (2008).

\bibitem{Kraeft06} W. D. Kraeft and M. Bonitz, J. Phys. Conf. Ser. \textbf{35%
}, 94 (2006).

\bibitem{Cioslowski08} J. Cioslowski and E. Grzebielucha, Phys. Rev E
\textbf{78}, 026416 (2008).

\bibitem{Wrighton09} J. Wrighton, J. W. Dufty, H. K\"{a}hlert, and M.
Bonitz. Phys. Rev. E \textbf{80}, 066405 (2009; arXiv:0909.0775.

\bibitem{Wrighton10} J. Wrighton, J. W. Dufty,  M.
Bonitz, and H. K\"{a}hlert. Contrib. Plasma Phys. \textbf{50}, 26
(2010); arXiv:0910.0076.

\bibitem{Thomson} J. J. Thomson, Philos. Mag. \textbf{7}, 237 (1904).

\bibitem{Dubin1999} D. Dubin and T. O'Neil. Rev. Mod. Phys. \textbf{71}, 87 (1999).

\bibitem{thomapp} C. Cecka, M. Bowick,  L. Giomi, A. Middleton, and K. Zielnicki, \emph{Thomson Problem--Points on a Sphere}, $<$http://thomson.phy.syr.edu/$>$

\end{thebibliography}
\end{document}